\documentclass[submitting]{nst}

\usepackage{subfigure,dcolumn}
\usepackage[T2A,T1]{fontenc}
\usepackage[russian,english]{babel}

\usepackage{threeparttable}

\makeatletter
\if@unpublished@  
  \let\linenumbers\relax  
\fi
\makeatother

\begin{document}

\title{Modelling instrumental response for neutron scattering experiments at CSNS}\thanks{This work was financially supported by the National Natural Science Foundation of China (Grant No.12075266), the National Key Research and Development Program of China (Grant No.2022YFA1604100) and Guangdong Basic and Applied Basic Research Fund (Guangdong-Dongguan Joint Fund) (Grant No.2023A1515140057).}

\author{Ni Yang}
\affiliation{Institute of High Energy Physics, Chinese Academy of Sciences, Beijing 100049, China}
\affiliation{Spallation Neutron Source Science Center, Dongguan 523803, China}
\affiliation{University of Chinese Academy of Sciences, Beijing 101408, China}

\author{Zi-Yi Pan}
\affiliation{Institute of High Energy Physics, Chinese Academy of Sciences, Beijing 100049, China}
\affiliation{Spallation Neutron Source Science Center, Dongguan 523803, China}

\author{Ming Tang}
\affiliation{Institute of High Energy Physics, Chinese Academy of Sciences, Beijing 100049, China}
\affiliation{Spallation Neutron Source Science Center, Dongguan 523803, China}

\author{Wen Yin}
\affiliation{Institute of High Energy Physics, Chinese Academy of Sciences, Beijing 100049, China}
\affiliation{Spallation Neutron Source Science Center, Dongguan 523803, China}

\author{Xiao-Xiao Cai}
\email[Corresponding author, Xiao-Xiao Cai, Address: Zhongziyuan Road, Dalang, Dongguan, Guangdong, China, Tel:+86 769-88932163,  E-mail:]{ caixx@ihep.ac.cn}
\affiliation{Institute of High Energy Physics, Chinese Academy of Sciences, Beijing 100049, China}
\affiliation{Spallation Neutron Source Science Center, Dongguan 523803, China}

\begin{abstract}
Thermal neutron total scattering experiments of light and heavy water were reproduced using the CSNS in-house Monte Carlo thermal neutron transport code, Prompt, with a focus on the instrumental detector response and the accurate derivation of thermal neutron scattering cross-sections.
In this work, a data reduction method is developed to process both the measured and simulated detector events for estimating angular, wavelength distributions, as well as angular differential cross sections.
The reduction results of simulations and experiments show a high degree of consistency. 
The prominent inelasticity signatures observed in the experiments can be accurately reproduced in simulations.
We discuss the  cause of the inelasticity effects, and demonstrate the elimination of such effects when the inelastic scattering process is taken into account in simulations.    
In addition, multiple scattering in samples is analysed and discussed.

\end{abstract}


\maketitle

\section{Introduction}
\label{sIntroduction}

It is vitally important to simulate thermal neutron scattering processes in the sample of a neutron instrument, as this enables the optimization of the instrument and, in addition, a clear understanding of the instrument’s response. By extension, this allows for more accurate interpretation of the measured data.
The thermal neutron scattering cross sections constitute a critical sub-library within the Evaluated Nuclear Data Files (ENDF), with major national evaluated nuclear data libraries worldwide including CENDL-3.2 ~\cite{ge2020cendl,zhang2022performance}, ENDF/B-VIII.0~\cite{brown2018endf},  JEFF-3.3~\cite{plompen2020joint} and JENDL-4.0~\cite{shibata2011jendl}, all incorporating these data as essential components. 
However, this sub-library only contains cross section data for approximately twenty materials, and such data are limited to fixed temperatures. Notably, these temperatures are typically above room temperature, as the sub-library was originally developed to meet the requirements of reactor applications.

To model crystalline materials, liquids, polymers and nano-materials at the precise experimental conditions at neutron scattering instruments, a thermal neutron scattering engine, NCrystal~\cite{Cai2020}, has been developed in collaboration by the researchers at the China Spallation Neutron Source (CSNS)~\cite{chen2016china,tang2021back,jie2009china,wei2009china,XU2021165642} and the European Spallation  Source  with robust numerical implementation~\cite{KITTELMANN2021108082,cai2019rejection} and detailed scattering processes~\cite{Cai_2017, Du2022}.


Monte Carlo ray-tracing technique, e.g. McStas~\cite{willendrup2021mcstas} 
, is widely used to optimize the performance of neutron scattering instruments~\cite{tang2023monte,wang2022design,han2019optimized}. 
Due to the simplified scattering physics and inherent linear chain approximation, it is often challenging to calculate the absolute scattering intensity. In addition, it is difficult to describe the non-linear physical arrangements in experiments, especially in the detector systems and samples/sample environments~\cite{lin2016mcvine}.

Recently, based on the physics in NCrystal, a new open-source Monte Carlo particle transport package, Prompt~\cite{PAN2024109004}, has been made available by the researchers from CSNS. 
It employs both Monte Carlo ray-tracing and transport techniques, allowing it to perform conventional ray-tracing simulations in neutron optics and to simulate detailed physics scattering processes in materials of arbitrary shape without the linear constraint. Such an advancement enables the instrumental response simulation at an absolute scale.

The neutron total scattering technique is a powerful approach for capturing scattering patterns over a wide range of momentum transfer, making it applicable for measuring both long-range and short-range order~\cite{xu2019physical,qin2021neutron,yang2025long}. 
As the energy transfer for each scattering event in total scattering experiments is unknown, it is challenging to correct inelasticity effects~\cite{soper2009inelasticity} on data, especially for light-element-rich materials.

The scope of this work focuses on leveraging all the physics in Prompt to model the interaction of the incident beam and directly compare it with experimental data, within which all principal distortion effects are taken into consideration. 
The method employed in this work slightly departs from the well-established analysis methods \cite{dawidowski2012data}, as it is not aimed at correcting distortions but at reproducing them using the software Prompt.
So that, such distortions can be better understood by the instrument users. 
Processes such as nuclear absorption, elastic scattering, inelastic scattering, coherent scattering, incoherent scattering, and multiple scattering in collimators, samples, and detectors are all considered in the simulations.
Section~\ref{sMethod} briefly introduces the time-of-flight neutron total scattering technique. Experiment and simulation setups are also given in this section, along with the unified data analysis method.
Section~\ref{sResults} compares the analysis results from experimental and simulated detector response.
Section~\ref{sDiscussion} analyses multiple scattering and inelasticity of the samples in Prompt.
Finally, the conclusions and prospect of this work are given in Section~\ref{sConclusion}.

\section{Method}
\label{sMethod}

\subsection{Data normalisation}
\label{ssDataNormalisation}

The implemented method for normalising the detector event count of the instrument is presented in this section.
  
\begin{figure}
\centering
\includegraphics[width=\linewidth]{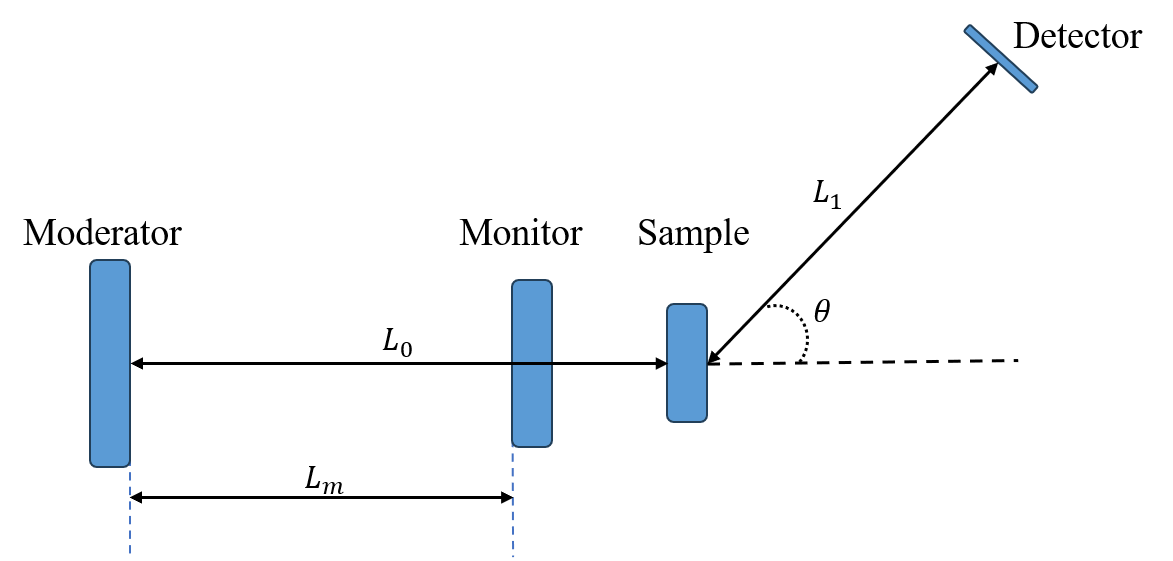}
\caption{Basic experimental setup of the time-of-flight neutron total scattering technique, showing the incident flight path distance $L_{0}$, the scattered flight path distance $L_{1}$, the moderator-to-monitor distance $L_{m}$ and the scattering angle $\theta$.}
\label{fbasic}
\end{figure}

The basic experimental setup~\cite{XU2021165642} of the time-of-flight (TOF) neutron total scattering technique is shown in Fig.~\ref{fbasic}.
Neutrons with wavelength $\lambda$ travel a distance $L_{0}$ from the moderator to the sample.
After interacting with the sample, the scattered neutrons with wavelength $\lambda^{\prime}$ traverse a distance $L_{1}$ to a detector placed at a scattering angle $\theta$.
The momentum transfer $\hbar Q$ and energy transfer $\hbar\omega$ are given by

\begin{equation} \label{eQ}
Q=2 \pi\left(\frac{1}{\lambda^{2}}+\frac{1}{{\lambda^{\prime}}^{2}}-\frac{2 \cos \theta}{\lambda \lambda^{\prime}}\right)^{\frac{1}{2}}
\end{equation}

\begin{equation} \label{eomega}
\omega=\frac{2 \pi^{2} \hbar}{m}\left(\frac{1}{\lambda^{2}}-\frac{1}{{\lambda^{\prime}}^{2}}\right)
\end{equation}
where $m$ is the mass of the neutron.
According to Eq.~(\ref{eomega}), $\lambda^{\prime}$ can be expressed as a function of $\lambda$ and $\omega$.
\begin{equation} \label{elambda'}
\lambda^{\prime}=\lambda\left(1-\frac{m \omega \lambda^{2}}{\pi h}\right)^{-\frac{1}{2}}
\end{equation}
The TOF for the detected neutrons is obtained by
\begin{equation} \label{etof1}
t=\frac{m}{h}\left(L_{0}\lambda+L_{1}\lambda^{\prime} \right)
\end{equation}
Inserting Eq.~(\ref{elambda'}) into Eq.~(\ref{etof1}), one can obtain
\begin{equation} \label{etof2}
t=\frac{m \lambda}{h}\left[L_{0}+L_{1}\left(1-\frac{m \omega \lambda^{2}}{ \pi h}\right)^{-\frac{1}{2}}\right]
\end{equation}

Under the neglect of sample absorption and multiple scattering contributions, the detected scattering event count rate per unit solid angle at TOF, $t$, at the scattering angle, $\theta$ can be calculated as ~\cite{powles1973analysis,palomino2011calibration}
\begin{equation} \label{ecountTOF}
c(t, \theta)=\int_{\substack{0 \\ t = \text{const}}}^{+\infty} d \lambda \Phi(\lambda) \varepsilon\left(\lambda^{\prime}, \theta\right) \frac{\lambda}{\lambda^{\prime}} S(Q, \omega) \left.\frac{\partial \omega}{\partial t} \right\rvert\, _{\lambda}
\end{equation}
Here $\Phi(\lambda)$ is the neutron fluence, $\varepsilon\left(\lambda^{\prime}, \theta\right)$ is the detection efficiency, $S(Q, \omega)$ is the scattering function~\cite{powles1973analysis}, and $\left.\frac{\partial \omega}{\partial t} \right\rvert\, _{\lambda}$ can be calculated using Eq.~(\ref{ecountTOF}).

For neutrons detected at TOF, $t$, at the scattering angle, $\theta$, the corresponding elastic wavelength $\lambda_{e}$ is defined as 
\begin{equation} \label{elambdaElastic}
\lambda_{e}=\frac{ht}{m\left(L_{0}+L_{1}\right)}
\end{equation}
The elastic momentum transfer $\hbar Q_{e}$ corresponding to $\lambda_{e}$ is obtained by inserting $\lambda=\lambda_{e}$ and $\lambda^{\prime}=\lambda_{e}$ into Eq.~(\ref{eQ}).
\begin{equation} \label{eQElastic}
 Q_{e}=\frac{2\pi \left[2\left(1-\cos{\theta}\right)\right]^{\frac{1}{2}}}{\lambda_{e}}
\end{equation}
Since $t$ can be represented as a function of $\lambda_{e}$ shown in Eq.~(\ref{elambdaElastic}), $c(t, \theta)$ in Eq.~(\ref{ecountTOF}) can be expressed as a function of $\lambda_{e}$.
\begin{equation} \label{ecountLambdaElastic1}
c(\lambda_{e}, \theta)=\int_{\substack{0 \\ t = \text{const}}}^{+\infty} d \lambda \Phi(\lambda) \varepsilon\left(\lambda^{\prime}, \theta\right) \frac{\lambda}{\lambda^{\prime}} S(Q, \omega) \left.\frac{\partial \omega}{\partial t} \right\rvert\, _{\lambda} \frac{dt}{d\lambda_{e}}
\end{equation}
Note that $c(\lambda_{e}, \theta)$ denotes the differential count per unit wavelength and per unit solid angle at $\lambda_{e}$ and  $\theta$.

According to Powles~\cite{powles1973analysis}, $S(Q, \omega)$ can be expanded in a Taylor series around $Q_{e}$.
\begin{equation} \label{eSqw}
S(Q, \omega)=S\left(Q_{e}, \omega\right)+\left.\frac{\partial S}{\partial Q}\right|_{Q=Q_{e}} \delta Q_{L}+\ldots
\end{equation}
Note that terms beyond the first order are neglected.
The deviation $\delta Q_{L}$ can be found in Eq.(4.17) of Ref.~\cite{powles1973analysis}.

When $S(Q, \omega)$ is truncated at the first term as  $S\left(Q_{e}, \omega\right)$, it is essentially the so-called elastic approximation. 
Although it introduces non-negligible errors for light-element-rich materials, where inelastic effects are pronounced,
it can still serve the study's objective to simulate the observed experimental data, including the consideration of inelastic effects.
Hence Eq.~(\ref{ecountLambdaElastic1}) becomes

\begin{equation} \label{ecountLambdaElastic2}
c(\lambda_{e}, \theta)=\int_{\substack{0 \\ t = \text{const}}}^{+\infty} d \lambda \Phi(\lambda) \varepsilon\left(\lambda^{\prime}, \theta\right) \frac{\lambda}{\lambda^{\prime}} S(Q_{e}, \omega) \left.\frac{\partial \omega}{\partial t} \right\rvert\, _{\lambda} \frac{dt}{d\lambda_{e}}
\end{equation}

To account for the distortion effects from absorption and multiple scattering processes, $c(\lambda_{e}, \theta)$ is scaled by the corresponding distortion factor $d(\lambda_{e}, \theta)$.
\begin{equation} \label{ecountLambdaElastic3}
\tilde{c}(\lambda_{e}, \theta)=c(\lambda_{e}, \theta)d(\lambda_{e}, \theta)
\end{equation}

In a TOF total scattering experiment, each detector event is characterised by the neutron TOF $t$ and detection location.
In principle, the detection locations should be calibrated by a separate Bragg diffraction measurement of a well-known sample.
Such calibration is not performed, as it is not expected to be critical for the water samples considered in the present work.

It is straight forward to obtain the scattering angle from the sample and detection locations.
As shown in Eq.~(\ref{elambdaElastic}), the elastic neutron wavelength $\lambda_{e}$ can be calculated from the TOF and the flight path distance. 
Hence, the detected scattering event count can be recorded in a two-dimensional histogram of $\lambda_{e}$ and $\theta$.
This histogram has $N_{\lambda_{e}}$ bins of equal width $\Delta \lambda_{e}$ for $\lambda_{e}$, and $N_{\theta}$ bins of equal width $\Delta \theta$ for $\theta$.
The center of the $j_{th}$ ($j=1,\ldots N_{\lambda_{e}}$) bin for $\lambda_{e}$ is $\lambda_{e,j}$, and the center of the $k_{th}$ ($k=1,\ldots N_{\theta}$) bin for scattering angle is $\theta_{k}$.
The detected scattering event count in the bin with the centers $\lambda_{e,j}$ and $\theta_{k}$ can be expressed as

\begin{equation} \label{eDeltaCount}
\tilde{C}(\lambda_{e,j}, \theta_{k})=  \tilde{c}(\lambda_{e,j}, \theta_{k})\Delta \lambda_{e}\Delta \Omega_{k}
\end{equation}
Here $\Delta \Omega_{k}$ is the corresponding solid angle coverage of the bin centered at $\theta_{k}$.
The differential count $\tilde{c}(\lambda_{e,j}, \theta_{k})$ can be estimated by
\begin{equation} \label{ecountDensity}
\tilde{c}(\lambda_{e,j}, \theta_{k})=\frac{\tilde{C}(\lambda_{e,j}, \theta_{k})}{\Delta \lambda_{e}\Delta \Omega_{k}}
\end{equation}

For the monitor located upstream of the sample at a distance $L_{m}$ (shown in Fig.~\ref{fbasic}), the observable of each detected event is TOF $t_{m}$. 
The corresponding wavelength $\lambda$ can be evaluated by 
\begin{equation} \label{elambdaMonitor}
\lambda=\frac{ht_{m}}{mL_{m}}
\end{equation}
Likewise, a histogram of $\lambda$ is applied to record the count of the monitor events.
The monitor event count in the bin with the center $\lambda_{j}$ is given by
\begin{equation} \label{eNumMonitor}
M(\lambda_{j}) = a_{m}\Phi(\lambda_{j})P_{m}(\lambda_{j})\Delta \lambda
\end{equation}
Here $a_{m}$ is the irradiated area of the monitor, and $P_{m}(\lambda_{j})$ is the probability that the monitor detects events at $\lambda_{j}$. 
Additionally, the values of $\lambda_{j}$ and $\Delta \lambda$ are equal to those of $\lambda_{e,j}$ and $\Delta \lambda_{e}$, respectively.
For the geometrically thin and weakly absorbing monitor, 
$P_{m}(\lambda_{j})$ can be regarded as proportional to the wavelength~\cite{farhi2009virtual}.

The wavelength spectrum normalised intensity, $I(\lambda_{e,j}, \theta_{k})$, is consequently computed as the ratio of the scattering event count $\tilde{C}(\lambda_{e,j}, \theta_{k})$ and the monitor event count $M(\lambda_{j})$.

\begin{equation} \label{eDeltaI}
I(\lambda_{e,j}, \theta_{k}) = \frac{\tilde{C}(\lambda_{e,j}, \theta_{k})}{M(\lambda_{j})}
\end{equation}

\subsection{Data analysis}
\label{ssDataAnalysis}

To obtain the absolute scattering intensity of a sample, two additional sets of data, for detector calibration and background subtraction, are needed in the analysis procedure.  

First, the normalised intensity of the empty container, $I_{c}(\lambda_{e,j}, \theta_{k})$, which includes the background signals, is subtracted from the normalised intensity of the sample with container, $I_{sc}(\lambda_{e,j}, \theta_{k})$.
This procedure is also applied to the normalised intensity of the vanadium with container, $I_{vc}(\lambda_{e,j}, \theta_{k})$.
\begin{equation} \label{eDeltaIsampleInExp}
 I_{s}(\lambda_{e,j}, \theta_{k})= I_{sc}(\lambda_{e,j}, \theta_{k})- I_{c}(\lambda_{e,j}, \theta_{k})
\end{equation}

\begin{equation} \label{eDeltaIvanadiumInExp}
 I_{v}(\lambda_{e,j}, \theta_{k})= I_{vc}(\lambda_{e,j}, \theta_{k})- I_{c}(\lambda_{e,j}, \theta_{k})
\end{equation}
Here the subscripts $s$, $v$ and $c$ respectively denote sample, vanadium and empty container.
The combinations $sc$ and $vc$ refer to sample with container and vanadium with container, respectively.

Next, aiming to remove the effects of the monitor and detection efficiency in $I_{s}(\lambda_{e,j}, \theta_{k})$, the calibrated intensity of the sample, $G(\lambda_{e,j}, \theta_{k})$, is given by
\begin{equation} \label{eDeltaG}
 G(\lambda_{e,j}, \theta_{k})= \frac{N_{v} I_{s}(\lambda_{e,j}, \theta_{k})}{N_{s} I_{v}(\lambda_{e,j}, \theta_{k})}
\end{equation}
where $N$ is the number of irradiated atoms in the material, which is calculated as the product of the corresponding number density and irradiated volume.

However, the calibration procedure described in Eq.~(\ref{eDeltaG}) also leads to the cancellation of the solid angle coverage $\Delta \Omega_{k}$, which is inherently included in $I_{s}(\lambda_{e,j}, \theta_{k})$ and introduced by Eq.~(\ref{eDeltaCount}).
Owing to the geometric configuration of the detectors, $\Delta \Omega_{k}$ exhibits angular dependence, stemming from the variable orientations and distances of the detectors relative to the sample.
In order to add the missing term $\Delta \Omega_{k}$ to $G(\lambda_{e,j}, \theta_{k})$, an idealised factor is calculated by the simulated data for an idealised isotropic scattering material, which has the same total scattering cross section, number density, and geometry as the vanadium.
In the simulation of the idealised material,
neutrons are scattered isotropically without energy exchange, absorption or multiple scatterings.
Furthermore, the detection efficiency of both the detectors and the monitor is 100\%.
Based on the Eq.~(\ref{eDeltaI}), the idealised factor, $ F(\lambda_{e,j}, \theta_{k})$, is defined as
\begin{equation} \label{eDeltaFactor}
  F(\lambda_{e,j}, \theta_{k})= \frac{I_{ideal}(\lambda_{e,j}, \theta_{k})}{N_{ideal} }
\end{equation}
where the subscript symbol $ideal$ represents the idealised material.
Eq.~(\ref{eDeltaCount}) and Eq.~(\ref{eDeltaI}) obviously demonstrate that $F(\lambda_{e,j}, \theta_{k})$  includes the term $\Delta \Omega_{k}$. 

The absolute scattering intensity, $G^{'}(\lambda_{e,j}, \theta_{k})$, is obtained by incorporating the corresponding solid angle coverage, as expressed by

\begin{equation} \label{eDeltaG2}
G^{'}(\lambda_{e,j}, \theta_{k}) = G(\lambda_{e,j}, \theta_{k}) F(\lambda_{e,j}, \theta_{k})
\end{equation}

According to Eq.~(\ref{eQElastic}), $F(\lambda_{e,j}, \theta_{k})$ and $G^{'}(\lambda_{e,j}, \theta_{k})$ can be reduced to $H_{F}(Q_{e})$ and $H_{G^{'}}(Q_{e})$, respectively.

Finally, the absolute scale of the derived differential scattering cross section (DDCS) for the sample is defined and computed as
\begin{equation} \label{eDCS}
\frac{d\sigma^{'}_{s} }{d\Omega} (Q_{e})= \frac{d\sigma_{v} }{d\Omega}\frac{ H_{G^{'}}(Q_{e}) }{H_{F}(Q_{e})}
\end{equation}
where the vanadium differential scattering cross section $d\sigma_{v}/ d\Omega=\sigma_{v}/4\pi$, and the vanadium scattering cross section, $\sigma_{v}$, is treated as a constant of \SI{5.1}{\barn}.

\subsection{Measurement}
\label{ssMeasurement}

Measurements of liquid light and heavy water were carried out on the  multi-physics instrument (MPI) at CSNS. 
These materials have been well-studied experimentally~\cite{kameda2003inelasticity,zeidler2012isotope,soper2013radial}.

The layout of MPI is shown in Ref.~\cite{XU2021165642}.
A decoupled water moderator with an effective surface of \SI{100}{\mm}(horizontal)$ \times $\SI{100}{\mm}(vertical) is located \SI{30}{\m} upstream of the sample.
A long-distance neutron guide, equipped with three bandwidth choppers, is placed between the moderator and a six-channel collimator.
This collimator restricts the neutron divergence on the sample to \SI{0.2}{\degree} in the horizontal direction.
A four-blade slit laid at \SI{0.5}{\m} upstream of the sample limits the beam size to \SI{10}{\mm}(h)$ \times $\SI{30}{\mm}(v).
After the slit, there is a TOF beam monitor \SI{0.457}{\m} in front of the sample.
A circle radial collimator, consisted of $\rm{B}_{4}\rm{C}$ blades, surrounds the sample.
The angular separation of this collimator blades is \SI{1.5}{\degree}.
Scattered neutrons through the collimator are detected by 28 detector modules covering a scattering angle range from \SI{31}{\degree} to \SI{170}{\degree}.
Each detector module is combined by eight adjacent $\rm^{3}{He}$ tubes with diameters of \SI{25.4}{\mm} at 20 bar.
Detector Tubes with two different lengths, i.e., \SI{300}{\mm} and \SI{500}{\mm}, are equipped in the instrument.
For every tube, each \SI{5}{\mm} segment along the length direction is defined as a pixel.
Details on the MPI detectors can be found in Table 1 of Ref.~\cite{XU2021165642}.

Total scattering experiments were performed on two liquid samples, light water and heavy water (Manufacturer: Aladdin, Product number: D113903), at room temperature.
The samples were placed in cylindrical vanadium containers with an inner radius of \SI{4.476}{\mm} and a thickness of \SI{0.29}{\mm}.
A vanadium stick with a radius of \SI{4.42}{\mm} and a height of \SI{30.04}{\mm} was chosen as the calibration sample.
An additional measurement on the empty container was also conducted for noise removal.
The details on the measurements are shown in Table~\ref{tableMeasurements}.

\begin{table}[htb]
	\centering                       
	\caption{Measurements carried out at the MPI instrument. Alone with the sample, a vanadium container is always placed at the sample position.}     
	\label{tableMeasurements} 
	
    \begin{tabular}{llllll}
    \toprule
     Sample & Volume & Density & Run ID & Time & Event rate     \\
      &(\SI{}{\centi\metre^3})  & (\SI{}{\gram\cdot\centi\metre^{-3}})  &  & (\SI{}{\second}) &   (\SI{}{events\cdot\second^{-1}})   \\
    \midrule
    $\rm{D}_{2}\rm{O}$& 0.90  & 1.105  & 0017672 & 7200 & 860149\\
    $\rm{H}_{2}\rm{O}$& 0.30  & 0.998  & 0017673  & 7200 & 848616  \\
    vanadium & 1.84 & 6.118  & 0017327 & 46950 & 804825 \\
    void & N/A & N/A & 0017262 & 57772 & 178997 \\
   
    \bottomrule
    \end{tabular}
\end{table}

\subsection{Prompt-based simulation}
\label{ssSimulation}

Light and heavy Water are well-studied theoretically~\cite{bellissent1995single,abe2014evaluation,damian2014cab}. 
The integral expressions for their observable cross sections in a time-of-flight total scattering setup are well established, as noted in Eq.(5) of~\cite{soper2009inelasticity} and Eq.(3) of~\cite{GRANADA1986223}.
In this work, the software Prompt~\cite{PAN2024109004} is utilized to perform the integration using a Monte Carlo approach.

\begin{figure}
\centering
\includegraphics[width=\linewidth]{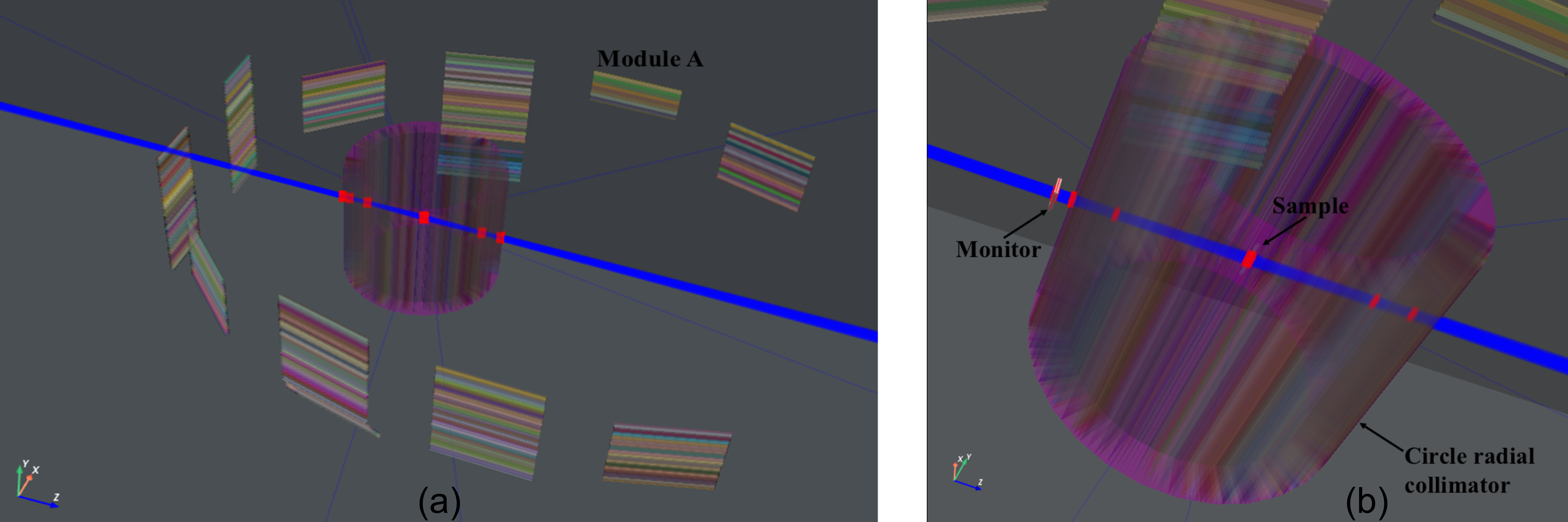}
 \caption{(a) Geometry visualisation of the simulated MPI in Prompt.
 (b) Zoom-in of the the simulated MPI.}
    \label{fMPIGeo}
\end{figure}

In order to reproduce the total scattering experiments using  Prompt, the components of MPI model are modelled and positioned according to the experimental setup described in Section~\ref{ssMeasurement}, including the monitor, circle radial collimator and all detector tubes. 
The sizes and positions of the moderator and slit are defined by a new particle gun \texttt{MPIGun}, the parameters of which are represented in Table~\ref{tableMPIGun}.
The horizontal divergency of \texttt{MPIGun} in simulations is set to \ang{0.2}.
\begin{table*}
\begin{threeparttable}[c]
\centering                       
\caption{Parameters available in \texttt{MPIGun} configuration strings. Parameters without default values are required. }     
\label{tableMPIGun} 	
\begin{tabular}{llll} 
    \toprule
    Parameter & Default & Description & Unit   \\ 

    \midrule
    \texttt{src\_w} &  & X-dimension (width) of particle source surface. & \si{\milli\meter} \\
    \texttt{src\_h} &  & Y-dimension (height) of particle source surface. & \si{\milli\meter} \\
    \texttt{src\_z} &  & Z-coordinate of particle source center in global reference system. & \si{\milli\meter} \\
    \texttt{slit\_w} &  & X-dimension (width) of slit surface. & \si{\milli\meter} \\
    \texttt{slit\_h} &  & Y-dimension (height) of slit surface. & \si{\milli\meter} \\
    \texttt{slit\_z} &  & Z-coordinate of slit center in global reference system. & \si{\milli\meter} \\
    
    \bottomrule
\end{tabular}
\end{threeparttable}
\end{table*}
Since the samples produce smooth angular distributions, the detailed geometries of the neutron guide and six-channel collimator, which are necessary for the accurate beam divergency, are not modelled.
However, as a result of this simplification, the neutron divergency in simulations is expected to differ slightly from that in experiments.
This can be later observed in Section~\ref{sResults} from the discrepancies between the simulated and measured vanadium Bragg peaks.
Fig.~\ref{fMPIGeo} shows the visualisation of MPI geometry in Prompt and 300 neutron trajectories.

Corresponding to the measurements in Table~\ref{tableMeasurements},
five simulations are carried out and listed in Table~\ref{tableSimulations}.
The simulations are conducted on an 88-core cluster composed of four Intel Xeon Gold 6238 processors, without employing any variance reduction methods. The geometries of the samples, the vanadium stick, and the vanadium container are made identical to those used in the experiments.
NCrystal~\cite{Cai2020} provides the cross section files of $\rm{D}_{2}\rm{O}$, $\rm{H}_{2}\rm{O}$ and vanadium, which are respectively  \texttt{LiquidHeavyWaterD2O\_T293.6K.ncmat}, \texttt{LiquidWaterH2O\_T293.6K.ncmat} and \texttt{V\_sg229.ncmat} in the standard library .
The cross sections for
$\rm{D}_{2}\rm{O}$ and $\rm{H}_{2}\rm{O}$ are evaluated by the CAB models~\cite{damian2014cab}.
\begin{table}
\begin{threeparttable}[c]
	\centering                       
	\caption{Simulations carried out in this work. Each run simulates \num{3e11} neutrons. Except for Run 5, the vanadium container is included in all the runs.}     
	\label{tableSimulations} 
	
    \begin{tabular}{lllll} 
    \toprule
    Run & Sample & Volume & Density & Time\tnote{1}  \\ 
    &  &(\SI{}{\centi\metre^3})  & (\SI{}{\gram\cdot\centi\metre^{-3}})   &  (CPU hours)   \\

    \midrule
    1 & $\rm{D}_{2}\rm{O}$ & 0.90  & 1.105  & 524.6 \\
    2 & $\rm{H}_{2}\rm{O}$ & 0.30  & 0.998  & 559.9 \\
    3 & vanadium  & 1.84 & 6.118    & 545.0 \\
    4 & void & N/A & N/A & 465.0\\
    5 & idealised scatter & 2.14& 6.118  & 414.1 \\
   
    \bottomrule
    \end{tabular}
\begin{tablenotes}
    \item [1] 
    The CPU time is calculated as the product of the number of cores and the hours each core is used.
    By applying the biasing technique~\cite{PAN2024109004} within Prompt, the computational efficiency can be significantly improved, as demonstrated in Section~\ref{ssMultiScat}.
\end{tablenotes}
\end{threeparttable}
\end{table}

New scorers named \texttt{WlAngle} in Prompt, directly producing the event count $\tilde{C}(\lambda_{e,j}, \theta_{k})$, are attached to the detectors.
The available parameters in \texttt{WlAngle} configuration strings are listed in Table~\ref{tableWlAngle}.
Aiming to consider the detection efficiency of the simulated detectors, only the scattering events absorbed by the $\rm^{3}{He}$ gas are recorded, i.e., \texttt{ptstate} of the scorers \texttt{WlAngle} are set to \texttt{ABSORB}.
The  monitor count $M(\lambda_{j})$ can be obtained by \texttt{WlSpectrum} scorers~\cite{PAN2024109004} attached to the monitors.
Since the information about monitor material is unclear, $P_{m}(\lambda_{j})$ for every simulated monitor is set to 100\%.
This means that all events incident on the monitors are counted, i.e., \texttt{ptstate} of the scorers \texttt{WlSpectrum}  are set to \texttt{ENTRY}. 

\begin{table*}
\begin{threeparttable}[c]
\centering                       
\caption{Parameters available in \texttt{WlAngle} configuration strings. Parameters without default values are required, the others are optional.}     
\label{tableWlAngle} 
	
\begin{tabular}{llll} 
    \toprule
    Parameter & Default & Description & Unit(Data Type)   \\ 

    \midrule
    \texttt{name} &  & Name of the histogram. &  str\\
    \texttt{sample\_pos} &  & Position vector of sample in global reference.
    & \si{\milli\meter}  \\
    \texttt{beam\_dir} &  & Direction vector of beam. & \si{\milli\meter}\\
    \texttt{dist} &  & Distance between neutron source and sample. & \si{\milli\meter}  \\
    \texttt{wl\_min} &  & Minimum wavelength of the histogram. & \si{\angstrom} \\ 
    \texttt{wl\_max} &  & Maximum wavelength of the histogram. & \si{\angstrom} \\
    \texttt{numbin\_wl} & 100 & Bin number of wavelength. & int \\
    \texttt{angle\_min} &  & Minimum angle of the histogram. & \si{\degree} \\ 
    \texttt{angle\_max} &  & Maximum angle of the histogram. & \si{\degree} \\
    \texttt{numbin\_angle} & 100 & Bin number of angle. & int \\
    \texttt{method} & 0  & Method for calculating wavelength\tnote{1}.  &  str\\
    \texttt{ptstate} & \texttt{ENTRY} & Particle tracing state, available options are: \texttt{SURFACE}, \texttt{ENTRY}, \texttt{ABSORB}, \texttt{PROPAGATE}, \texttt{EXIT}. & str\\
    \texttt{msname} & `` '' & Name of the \texttt{MultiScat} scorer attached to the scatterer\tnote{2}.   & str \\
    \texttt{scatnum} & -2 & Scattering number to be counted\tnote{3}.   & int \\
    
    \bottomrule
\end{tabular}
\begin{tablenotes}
    \item [1] The default value $0$ implies the method considering energy transfer, while a value of $1$ means the method based on the assumption that neutrons do not change energy in the scattering.
    \item [2] The scattering number of each event is counted by this \texttt{MultiScat} scorer.
    \item [3] Should be an integer greater than or equal to -2. 
    The default value implies that the histogram accumulates events regardless of the scattering number.
    The value -1 implies the accumulation on the events that do not enter the scatterer attached by the \texttt{MultiScat} scorer specified by \texttt{msname}.
    Other values enable the accumulation on the specified scattering number. 
\end{tablenotes}
\end{threeparttable}
\end{table*}

\section{Results}
\label{sResults}

\begin{figure}
\centering
\includegraphics[width=\linewidth]{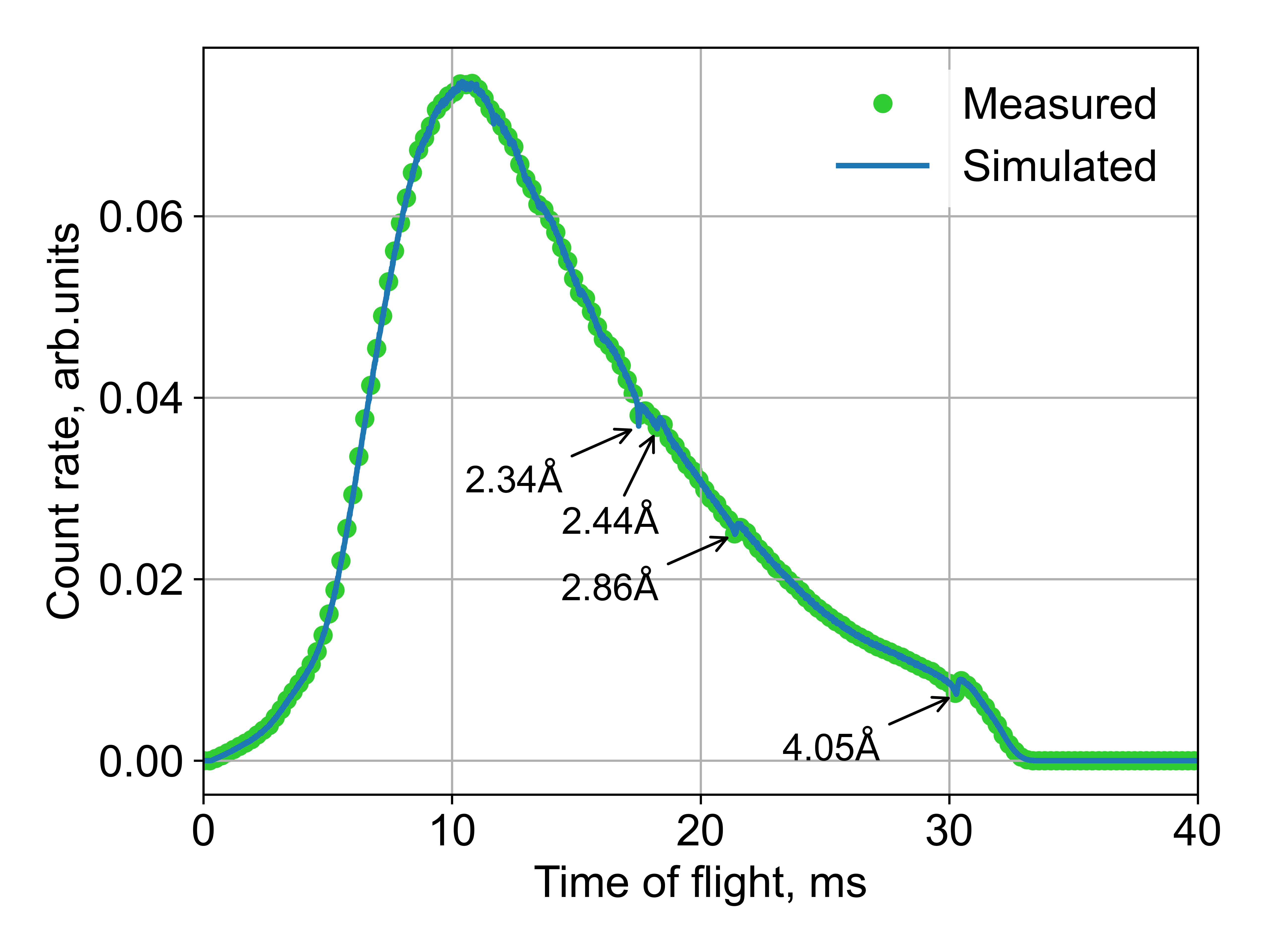}
\caption{TOF count rate for  monitors in the measurement (dots) and simulation (solid line).
Since the distance from the moderator to the monitor is \SI{29.543}{\m}, the corresponding wavelengths of the four sharp dips observed in both results are respectively $\SI{2.34}{\angstrom}$,  $\SI{2.44}{\angstrom}$, $\SI{2.86}{\angstrom}$ and $\SI{4.05}{\angstrom}$. 
}\label{fmonitor_TOF}
\end{figure}

Fig.~\ref{fmonitor_TOF} presents the distribution of neutron fluence at the initial wavelength for simulated neutrons, derived from the experimental TOF data of the monitor according to Eq.~(\ref{eNumMonitor}).
The curve of the TOF count rate for the simulated monitor, which is filled with a tiny amount of $\rm^{3}{He}$ gas, overlaps with the measured data.
This confirms that the incident wavelength is correctly modelled.

\begin{figure}
\centering
\includegraphics[width=\linewidth]{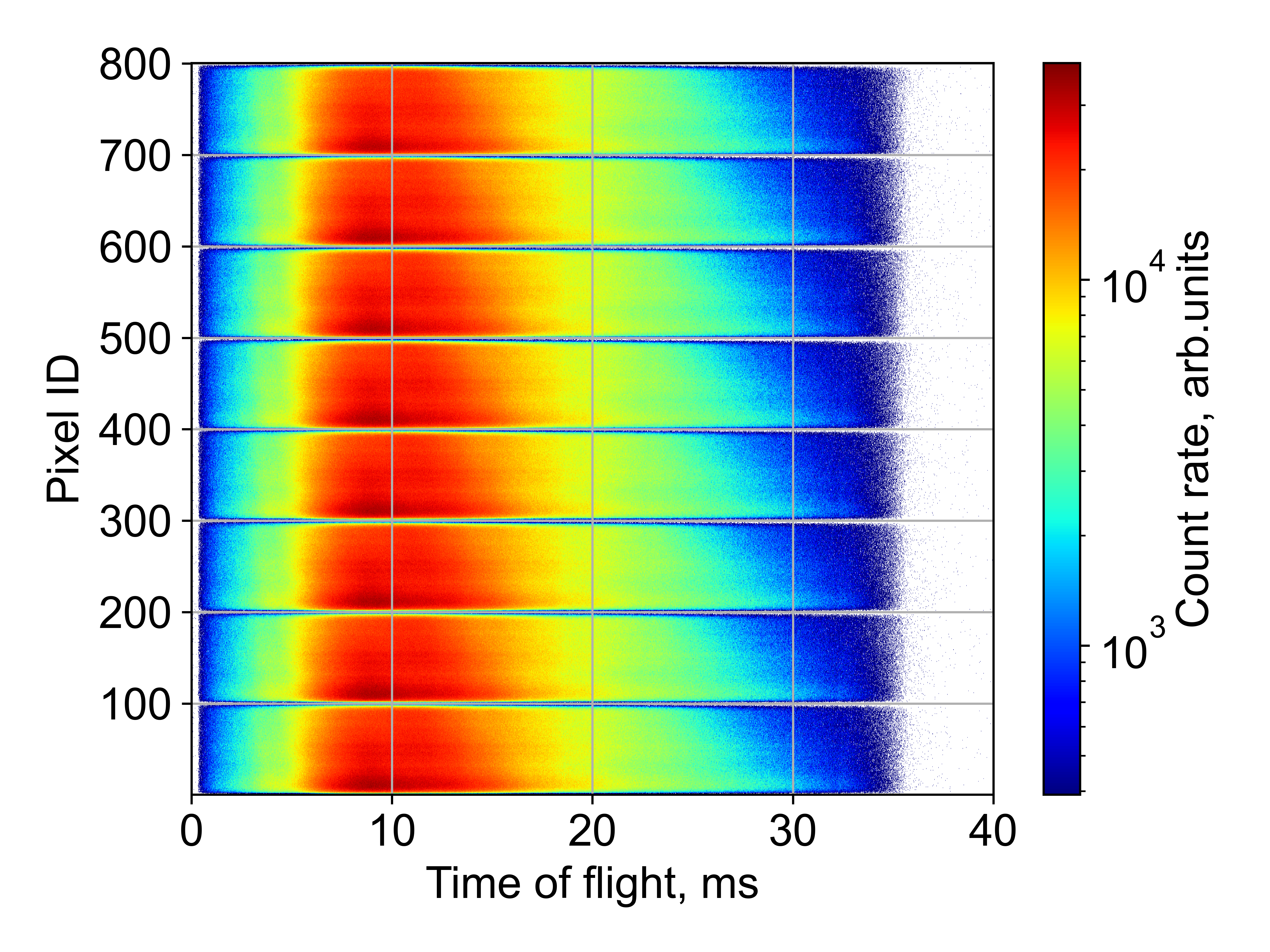}
\caption{TOF count rate of all 800 pixels in detector module A for the measurement of $\rm{D}_{2}\rm{O}$ sample with container.
}\label{ftof_pixel_hwcExp}
\end{figure}

Fig.~\ref{ftof_pixel_hwcExp} shows the TOF data of $\rm{D}_{2}\rm{O}$ measurement (Run 0017672), which are collected by all the pixels of detector module A (see Fig.~\ref{fMPIGeo}(a)). 
Module A consists of eight detector tubes, and each tube has 100 pixels.
It can be observed that the TOF data exhibit a periodic pattern every 100 pixels, suggesting that the response of the adjacent detector tubes is similar.

\begin{figure}
\centering
\includegraphics[width=\linewidth]{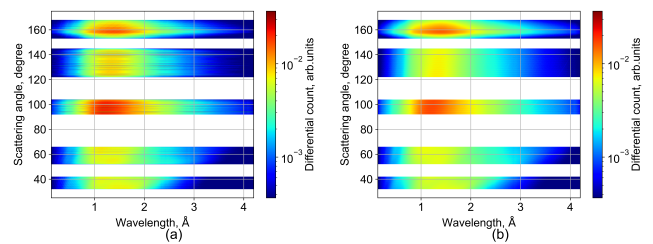}
\caption{Differential count (Eq.~(\ref{ecountDensity})) in wavelength and scattering angle for $\rm{D}_{2}\rm{O}$ sample with container. 
(a) Measured data and (b) the corresponding  simulated data. }
\label{feventDistri_hwcExpSim}
\end{figure}

According to Eq.~(\ref{eDeltaCount}) and Eq.~(\ref{ecountDensity}), the TOF data from all detector modules for the $\rm{D}_{2}\rm{O}$ measurement and simulation are converted to the differential count in wavelength and scattering angle, as shown in Fig.~\ref{feventDistri_hwcExpSim}. 
The blank regions observed in both the measured and simulated results are due to the limited angular coverage of the detectors.
The scattering angle ranges of the experimental and simulated data appear consistent, verifying that the detectors are placed consistently in the experiments and simulations.
Moreover, the simulated data are broadly similar to the experimental data, although the latter exhibit significant fluctuations.
The discrepancies can be attributed to the deviations in the experimental detection efficiency from the nominal values.

\begin{figure}
\centering
\includegraphics[width=\linewidth]{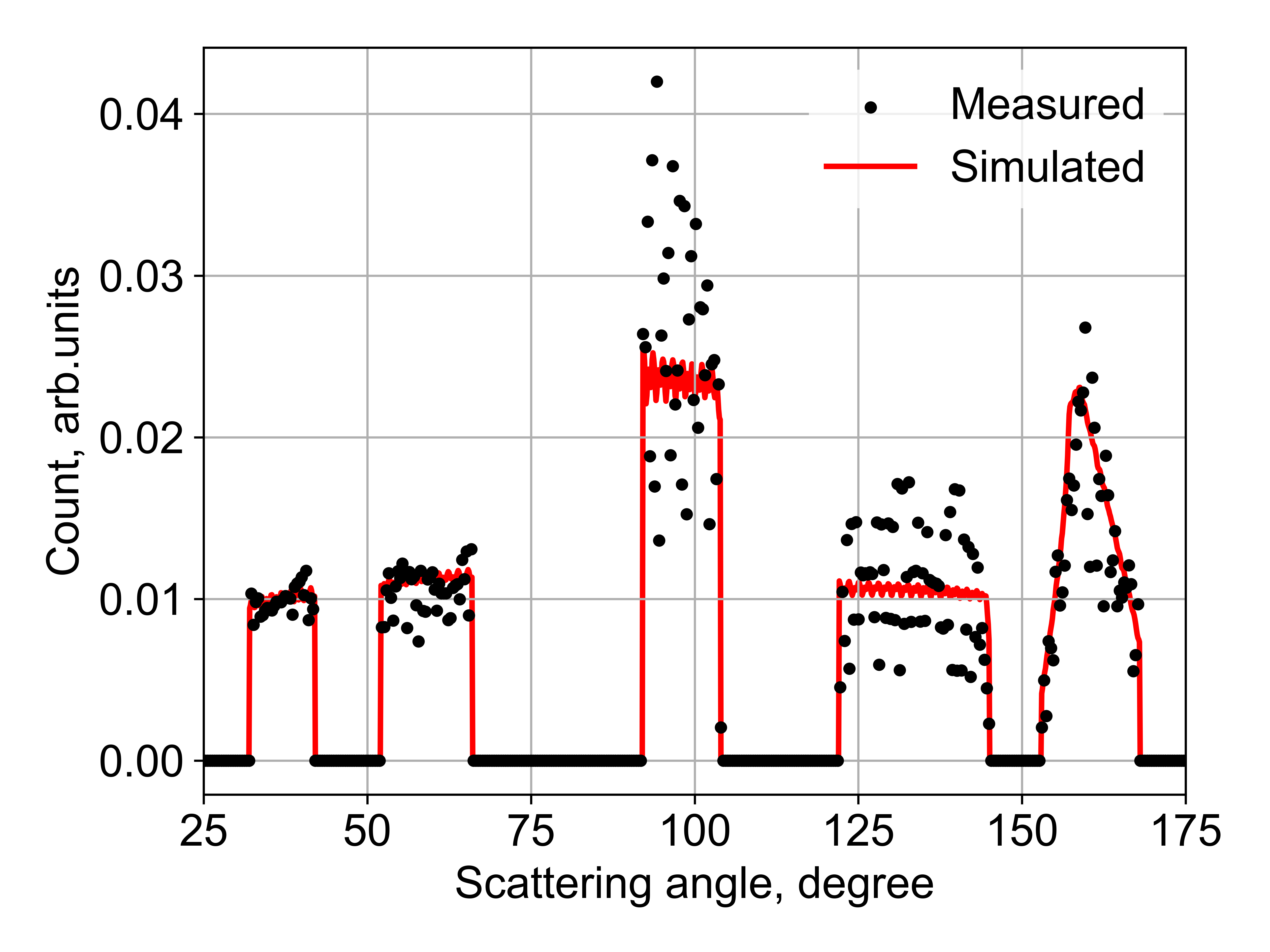}
\caption{Count in scattering angle for the measurement (dots) and simulation (solid line) of $\rm{D}_{2}\rm{O}$ sample with container, calculated by integrating the data from Fig.~\ref{feventDistri_hwcExpSim} over the wavelength.  
}\label{finteg_eventDistri_hwcExpSim}
\end{figure}

Fig.~\ref{finteg_eventDistri_hwcExpSim} compares the measured and simulated count in scattering angle for $\rm{D}_{2}\rm{O}$, which is obtained by integrating the data from Fig.~\ref{feventDistri_hwcExpSim} over the wavelength.
Despite the noticeable fluctuations in the measured data,
the shapes of the two results show good agreement.
Since the statistical error of the measurement is smaller than that of the simulation, the measured fluctuations are expected to be caused by the non-uniform detection efficiency.

\begin{figure}
\centering
\includegraphics[width=\linewidth]{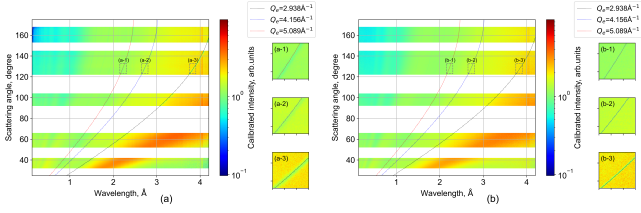}
\caption{Calibrated intensities (Eq.~(\ref{eDeltaG})) of $\rm{D}_{2}\rm{O}$.
    (a) Measured data and (b) the simulated data. 
    The black, blue, and red dotted lines represent $Q_{e}$=\SI{2.938}{\per\angstrom}, \SI{4.156}{\per\angstrom}, and \SI{5.089}{\per\angstrom}, respectively. 
    Subplots (a-1), (a-2), and (a-3) provide the magnified views of the corresponding sections in Figure (a) to highlight the valleys caused by the Bragg peaks of the vanadium.
    Similarly, subplots (b-1), (b-2), and (b-3)  demonstrate the simulated valleys in Figure (b).
    } \label{fG_hwExpSim}
\end{figure}

\begin{figure}
\centering
\includegraphics[width=\linewidth]{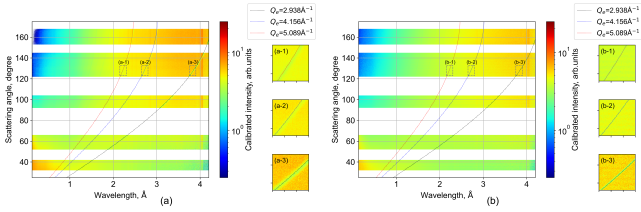}
\caption{Calibrated intensities (Eq.~(\ref{eDeltaG})) of $\rm{H}_{2}\rm{O}$.
    (a) Measured data and (b) the simulated data. 
    The black, blue, and red dotted lines represent $Q_{e}$=\SI{2.938}{\per\angstrom}, \SI{4.156}{\per\angstrom}, and \SI{5.089}{\per\angstrom}, respectively. 
    Subplots (a-1), (a-2), and (a-3) provide the magnified views of the corresponding sections in Figure (a) to highlight the valleys caused by the Bragg peaks of the vanadium.
    Similarly, subplots (b-1), (b-2), and (b-3)  demonstrate the simulated valleys in Figure (b).
    } \label{fG_lwExpSim}
\end{figure}

The measured and simulated calibrated intensities of $\rm{D}_{2}\rm{O}$, calculated by Eq.~(\ref{eDeltaG}), are respectively presented in Fig.~\ref{fG_hwExpSim}(a) and Fig.~\ref{fG_hwExpSim}(b).
It is noted that the simulated result is generally in good agreement with the measured one.
The valleys corresponding to the vanadium Bragg peaks~\cite{zemann1965crystal} of the (110) plane at $Q_{e}=\SI{2.938}{\per\angstrom}$, the (200) plane at $Q_{e}=\SI{4.156}{\per\angstrom}$, and the (211) plane at $Q_{e}=\SI{5.089}{\per\angstrom}$ are clearly visible.
These valleys result from the vanadium in the calibration process, as mentioned in Eq.~(\ref{eDeltaG}).
However, the measured valleys have a poorer resolution than the simulated ones.
This indicates the tiny differences between the experimental and simulated divergence, arising from the simplification of the simulated source term.
Additionally, the measured valleys are slightly off the expected trajectories as the pixel locations in the analysis are not calibrated to a standard silicon sample.
The above-mentioned behaviors can also be observed in the calibrated intensities of $\rm{H}_{2}\rm{O}$ from measurements and simulations (see Fig.~\ref{fG_lwExpSim}). 

\begin{figure}
\centering
\includegraphics[width=\linewidth]{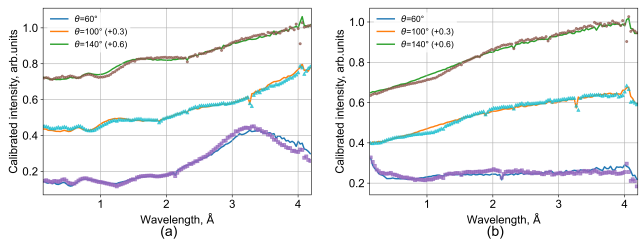}
\caption{(a) The measured and simulated calibrated intensities (Eq.~(\ref{eDeltaG})) of $\rm{D}_{2}\rm{O}$ at scattering angles of \SI{60}{\degree}, \SI{100}{\degree}, and \SI{140}{\degree}. 
(b)The corresponding measured and simulated results for $\rm{H}_{2}\rm{O}$.
The solid lines are obtained from the simulations, while squares, triangles and dots are respectively the measured data at \SI{60}{\degree}, \SI{100}{\degree}, and \SI{140}{\degree}.   
    } \label{fG_ExpSim_cut}
\end{figure}

The measured and simulated calibrated intensities of $\rm{D}_{2}\rm{O}$ at scattering angles of \SI{60}{\degree}, \SI{100}{\degree}, and \SI{140}{\degree} are extracted from Fig.~\ref{fG_hwExpSim} and shown in Fig.~\ref{fG_ExpSim_cut}(a).
The structures observed in the measured data are well captured by the simulated curves.
Fig.~\ref{fG_ExpSim_cut}(b) presents the corresponding calibrated intensities of $\rm{H}_{2}\rm{O}$ extracted from Fig.~\ref{fG_lwExpSim}.
Similarly, the simulated results are consistent with the measured data when the wavelength exceeds \SI{1.5}{\angstrom}.
However, the gentle troughs observed within the \SI{0.8}{\angstrom} to \SI{1.5}{\angstrom} wavelength range in the experimental results for the three different scattering angles are not reproduced by the simulations.
Additionally, the peak at \SI{0.46}{\angstrom} in the experimental data for the scattering angle of \SI{60}{\degree} is also not captured.
These discrepancies originate from the absence of the coherent scattering processes of $\rm{H}$ and $\rm{O}$ in the $\rm{H}_{2}\rm{O}$ simulation~\cite{damian2014cab}.

\begin{figure}
\centering
\includegraphics[width=\linewidth]{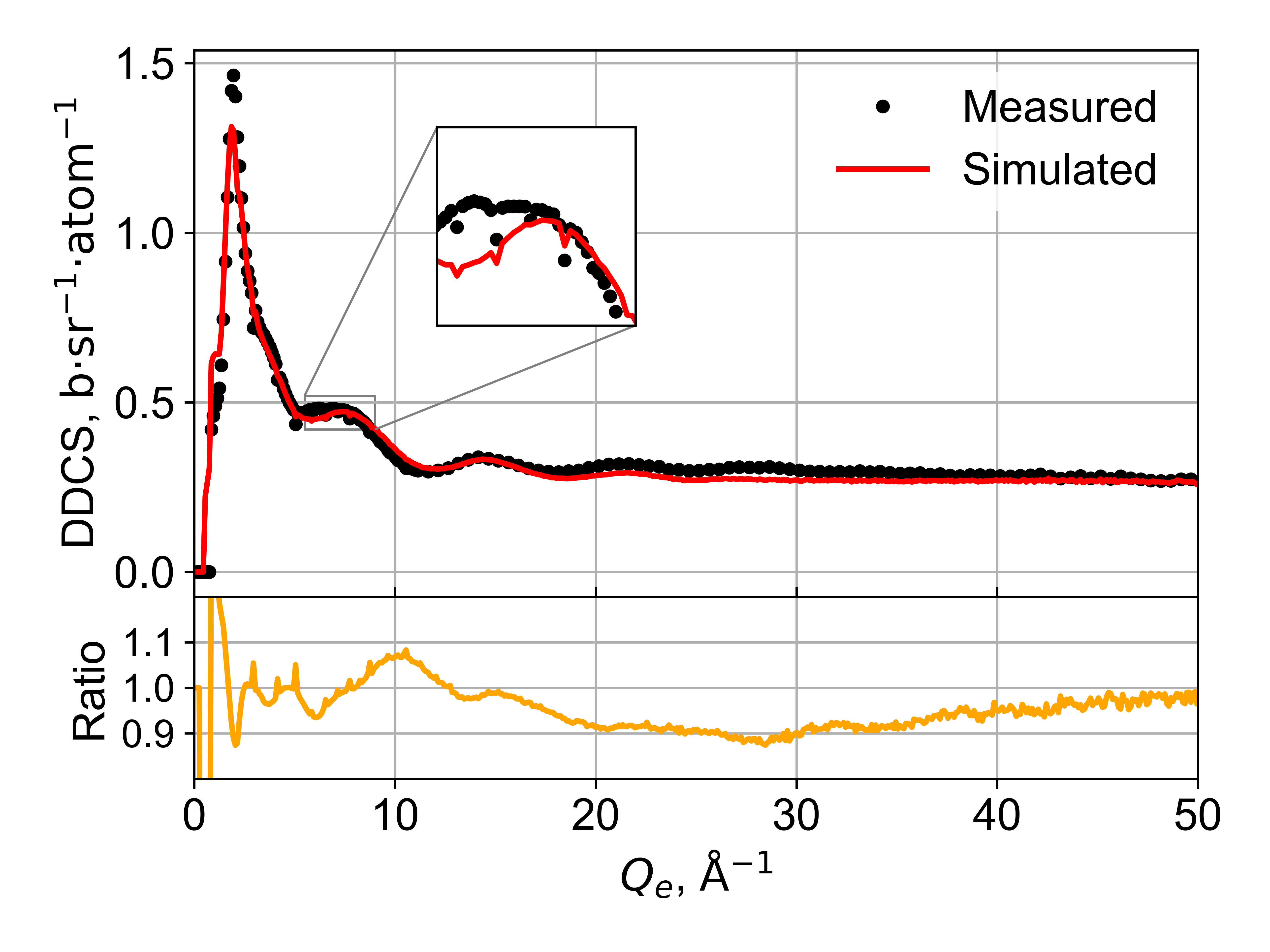}
\caption{Derived differential cross sections (Eq.~(\ref{eDCS})) for $\rm{D}_{2}\rm{O}$ obtained from measurements (dots) and simulations (red solid line).
Orange solid line shows the ratio of simulated result to the measured one.
}\label{fDDCS_hw}
\end{figure}

\begin{figure}
\centering
\includegraphics[width=\linewidth]{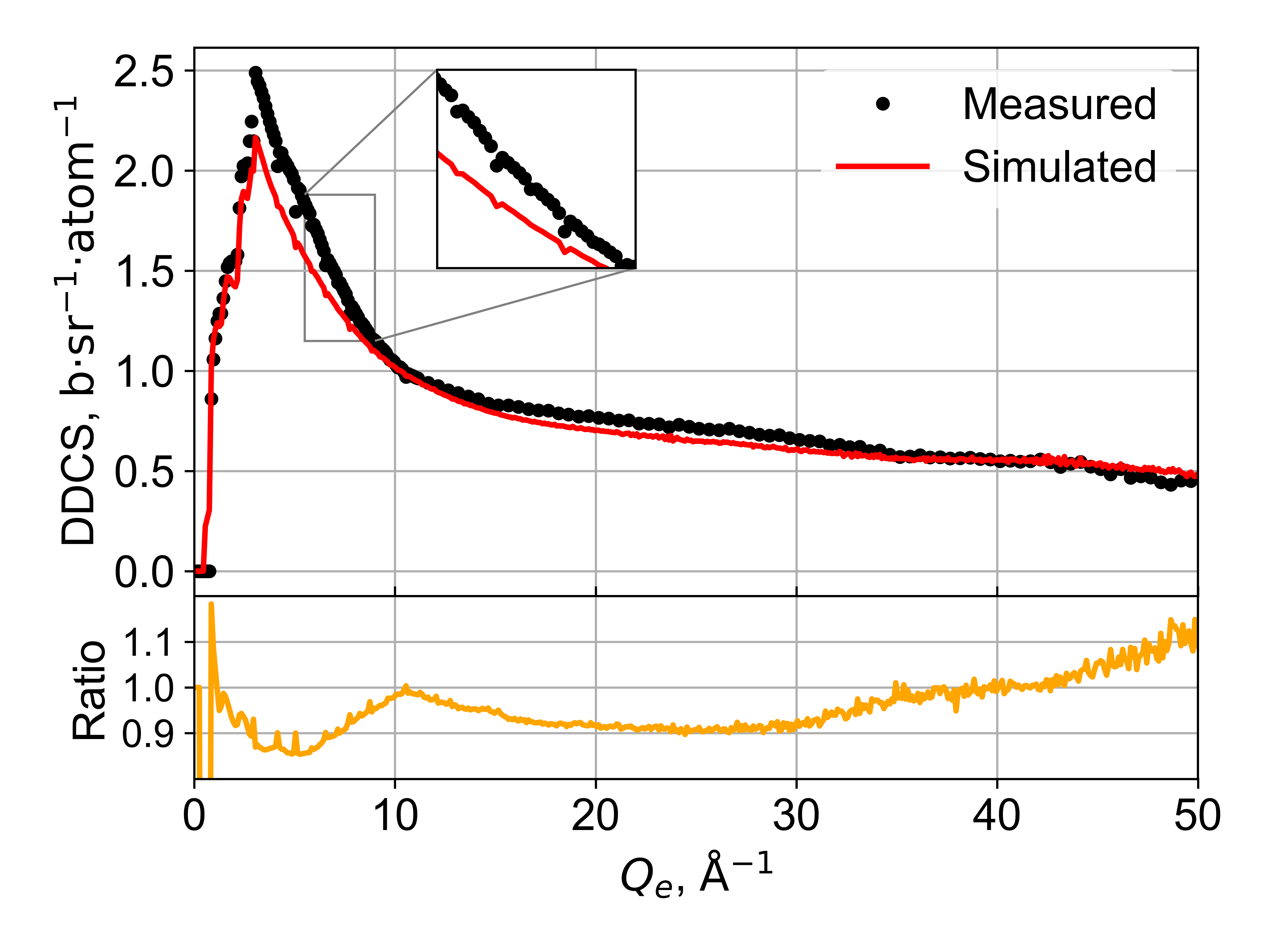}
\caption{Derived differential cross sections (Eq.~(\ref{eDCS})) for $\rm{H}_{2}\rm{O}$ obtained from measurements (dots) and simulations (red solid line).
Orange solid line shows the ratio of simulated result to the measured one.
}\label{fDDCS_lw}
\end{figure}

In order to assess the consistency of experiments and simulations in terms of $Q_{e}$, the derived differential cross sections for $\rm{D}_{2}\rm{O}$ and $\rm{H}_{2}\rm{O}$ at absolute scale are obtained by applying Eq.~(\ref{eDCS}) to the measured and simulated data collected from all detector modules.
The results for $\rm{D}_{2}\rm{O}$ are shown in Fig.~\ref{fDDCS_hw} and for $\rm{H}_{2}\rm{O}$ in Fig.~\ref{fDDCS_lw}.
Apart from the small discrepancies in the peaks at \SI{7.4}{\per\angstrom} for $\rm{D}_{2}\rm{O}$ and the peaks at \SI{1.7}{\per\angstrom} for $\rm{H}_{2}\rm{O}$, the structures of the measured curve are well captured by simulations, including the dips introduced by the Bragg effect of the vanadium, as shown in the zoom-in insets of Fig.~\ref{fDDCS_hw} and Fig.~\ref{fDDCS_lw}. 
Since scattering angles below \SI{31}{\degree} are not covered by the detectors, as illustrated in Fig.~\ref{fMPIGeo}(a), drops in the low $Q_{e}$ zones are present in both the $\rm{D}_{2}\rm{O}$ and $\rm{H}_{2}\rm{O}$ results.
At absolute scales, the disagreements are less than 15\% when $Q_{e}$ is greater than \SI{1.5}{\per\angstrom}.
Sharp structures on the ratio curves for $Q_{e}$ lower than \SI{1.5}{\per\angstrom} are induced by statistical errors.
The shapes of the ratio curves for $\rm{D}_{2}\rm{O}$ and $\rm{H}_{2}\rm{O}$ are broadly similar, indicating that the sources of the discrepancies for both samples are consistent.
Notice that the scattering cross section for $\rm{D}_{2}\rm{O}$~\cite{damian2014cab} is generated under the Sk\"old approximation~\cite{skold1967small} using the refined static structure factor~\cite{soper2013radial}. 
Consequently, slight discrepancies due to the refinement procedure are expected.

\section{Discussion}
\label{sDiscussion}

\subsection{Multiple scattering}
\label{ssMultiScat}

The complete evaluation of the distortion factor, $d(\lambda_{e}, \theta)$ in Eq.~(\ref{ecountLambdaElastic3}) is beyond the scope of this manuscript, and will be discussed elsewhere in the near future. In this section, we implement the biasing technique, a variance reduction technique within Prompt, to efficiently evaluate the contributions from multiple scattering.  

\begin{figure}
\centering
\includegraphics[width=\linewidth]{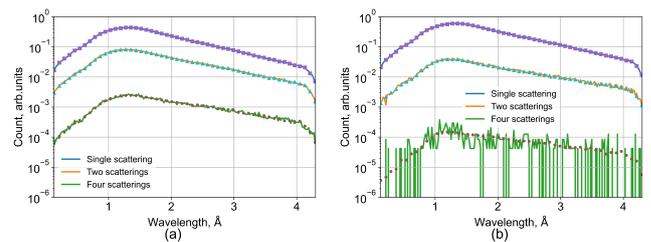}
 \caption{Count in wavelength for different scattering numbers, calculated by integrating the differential count in Eq.~(\ref{ecountDensity}) over the scattering angle.
    (a) The results obtained from non-biased and biased simulations of $\rm{D}_{2}\rm{O}$ cylindrical sample with container, and (b) the corresponding results from the simulations of $\rm{D}_{2}\rm{O}$ slab.
The incident neutron numbers for all the simulations in (a) and (b) are $10^{9}$.
Solid lines are the non-biased results, while squares, triangles and dots are respectively the biased results of single scattering, two scatterings and four scatterings.}
    \label{fIntegral_bias}
\end{figure}

The single and multiple scattering contributions are extracted by counting the number of segmentation of neutron trajectories in the sample.
Fig.~\ref{fIntegral_bias}(a) presents the the single, two and four contributions to the count of all detectors for the $\rm{D}_{2}\rm{O}$ simulation, as mentioned in Table~\ref{tableSimulations}.
Compared with the contributions of higher scattering numbers, the influence caused by two scatterings is the most significant.
It is evaluated that the event count of two scatterings is 19.2\% of that of single scattering.
Since the curve shapes for different scattering numbers are similar, there is little difference between the shapes of the results for single scatterings and all scatterings.
Considering the sample thickness of \SI{8.952}{\mm}, the strong multiple scattering noise results from this excessive thickness.
To test the impact of thickness on multiple scattering, the contributions of multiple scattering for a $\rm{D}_{2}\rm{O}$ slab with a thickness of \SI{1}{\mm} are calculated and shown
in Fig.~\ref{fIntegral_bias}(b).
Not surprisingly, the ratio of the event counts of two scatterings and single scattering is greatly reduced to 6.8\%.

Fig.~\ref{fIntegral_bias}(a) and Fig.~\ref{fIntegral_bias}(b) also show the results of single and multiple scattering contributions calculated by the cross section biasing technique.
The results of the non-biased runs show that the statistical qualities for different scattering numbers get worse as the scattering number increases.
However, the biased curves for high scattering numbers exhibit statistically superior performance compared to their non-biased counterparts.
This observation suggests that the biasing technique can significantly enhance the computational efficiency, especially when analysing physical processes with large variances.

\subsection{Inelasticity signature}

For strong inelastic samples, the assumption that neutrons do not exchange energy with the sample, which is made in Eq.~(\ref{ecountLambdaElastic2}), no longer holds true.
The effects of inelasticity arise from the deviation of the wavelength spectrum at the detectors, following neutron interactions with the sample, from the original incident spectrum.

\begin{figure}
\centering
\includegraphics[width=\linewidth]{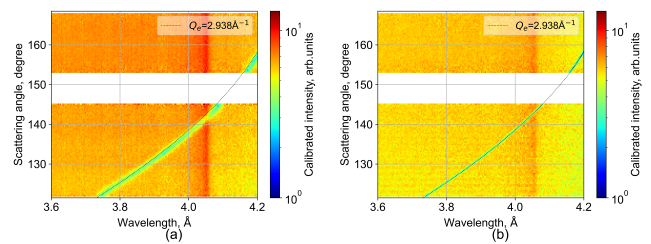}
 \caption{(a) The inelasticity signature of $\rm{H}_{2}\rm{O}$ in the measured data, and (b) the corresponding simulated result with the elastic approximation in the \texttt{WlAngle} scorer.
 (a) and (b) are zoomed-in sections of the wavelength range 
$\SI{3.6}{\angstrom}\sim\SI{4.2}{\angstrom}$ 
 and the scattering angle range 
 $\SI{121.5}{\degree}\sim\SI{168.5}{\degree}$
 from Fig.~\ref{fG_lwExpSim}(a) and Fig.~\ref{fG_lwExpSim}(b), respectively.
 }\label{fG_lwExpSim_zoomin}
\end{figure}

\begin{figure}
\centering
\includegraphics[width=\linewidth]{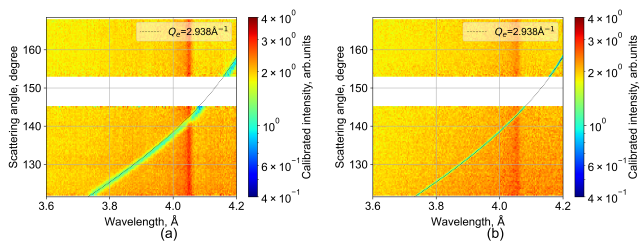}
 \caption{(a) The inelasticity signature of $\rm{D}_{2}\rm{O}$ in the measured data, and (b) the corresponding simulated result with the elastic approximation in the \texttt{WlAngle} scorer.
  (a) and (b) are zoomed-in sections of the wavelength range 
$\SI{3.6}{\angstrom}\sim\SI{4.2}{\angstrom}$ 
 and the scattering angle range 
 $\SI{121.5}{\degree}\sim\SI{168.5}{\degree}$
 from Fig.~\ref{fG_hwExpSim}(a) and Fig.~\ref{fG_hwExpSim}(b), respectively.
 }\label{fG_hwExpSim_zoomin}
\end{figure}

Several unexpected peak structures, which are independent of wavelength and therefore exclude the possibility of Bragg diffraction, are observed in both the experimental calibrated intensities and the corresponding simulated results (see Fig.~\ref{fG_hwExpSim} and Fig.~\ref{fG_lwExpSim}) . 
The locations of the peaks are coincidentally consistent with the small dip structures in the incident spectrum. 
Those sharp dips are caused by the coherent elastic scattering in the metallic materials, most likely aluminum, along the beam path, as shown in Fig.~\ref{fmonitor_TOF}.
For better visualisation purpose, these peak structures are zoomed in within Fig.~\ref{fG_lwExpSim_zoomin} for light water and Fig.~\ref{fG_hwExpSim_zoomin} for heavy water, focusing on the wavelength range from
$\SI{3.6}{\angstrom}$ to $\SI{4.2}{\angstrom}$.

\begin{figure}
\centering
\includegraphics[width=\linewidth]{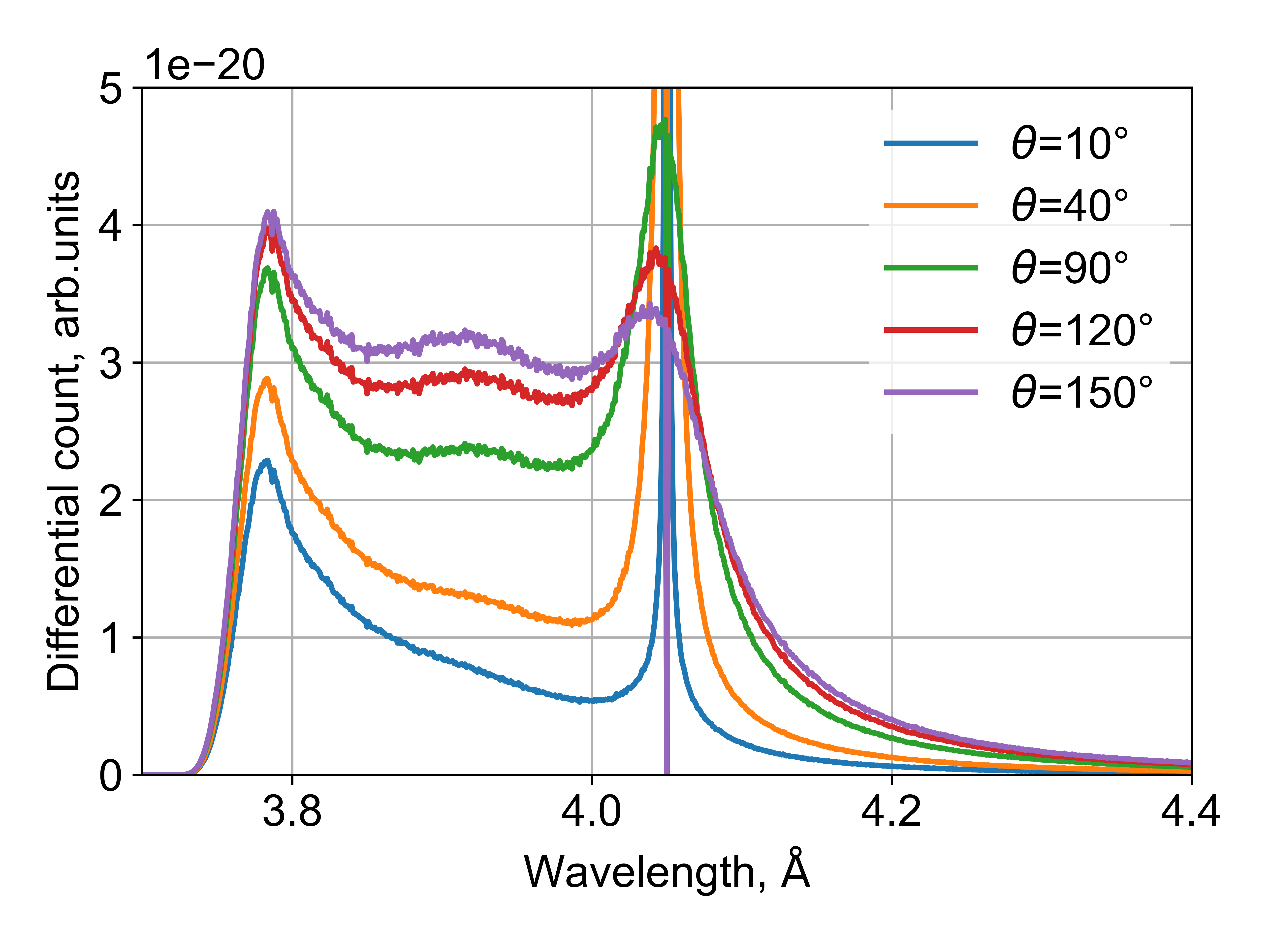}
 \caption{Normalised differential count as a function of wavelength for scattering angles $\theta$=\SI{10}{\degree}, \SI{40}{\degree}, \SI{90}{\degree}, \SI{120}{\degree} and \SI{150}{\degree}. 
 }\label{f_wl_newSim}
\end{figure}

\begin{figure}
\centering
\includegraphics[width=\linewidth]{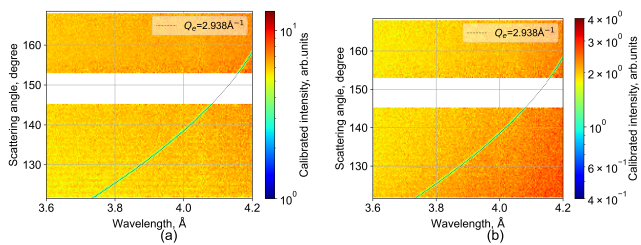}
 \caption{(a) The simulated result for $\rm{H}_{2}\rm{O}$ with the inelastic treatment in the \texttt{WlAngle} scorer, and (b) the corresponding simulated result for $\rm{D}_{2}\rm{O}$.
 }\label{fG_ine_zoomin}
\end{figure}

To further investigate the inelasticity of the monoenergetic \SI{4.05}{\angstrom} incident neutrons, 
an additional simulation for $\rm{H}_{2}\rm{O}$ is performed.
In this simulation, a small sample is surrounded by a perfect spherical detector with 100\% detection efficiency.
Fig.~\ref{f_wl_newSim} shows the normalised differential count,  $\tilde{c}_{n}(\lambda_{e,j},\theta_{k})=\tilde{c}(\lambda_{e,j}, \theta_{k})/[M(\lambda_{j})/\Delta\lambda]$, at scattering angles of \SI{10}{\degree}, \SI{40}{\degree}, \SI{90}{\degree}, \SI{120}{\degree} and \SI{150}{\degree}.
The ratios of the accelerated neutrons after scattering, with wavelengths shorter than the incident wavelength, are tabulated in Table~\ref{table_wl_newSim}.

For all the five scattering angles, over 76\% of the scattering events exhibit a pronounced shift toward shorter wavelengths due to
inelastic scattering with hydrogen. 
This preferential energy gain shifts the wavelengths of scattered neutrons, causing the detector spectrum itself to be shifted and distorted compared to the spectrum from an almost elastic vanadium calibration sample. 
Crucially, when the raw sample
spectrum, containing this inelastic distortion, is divided by the monitor spectrum, as indicated in the expression of $\tilde{c}_{n}(\lambda_{e,j},\theta_{k})=\tilde{c}(\lambda_{e,j}, \theta_{k})/[M(\lambda_{j})/\Delta\lambda]$, 
the resulting normalised Bragg peak positions, including the \SI{4.05}{\angstrom} edge, appear shifted to shorter
wavelengths.
This phenomenon directly leads to the sharp dips at \SI{4.05}{\angstrom} observed in Fig.~\ref{f_wl_newSim}.

Therefore, it is evident that the inelasticity alone introduces the unexpected peaks at the locations of the shape dips in the incident spectrum. To prove that, two new simulations of $\rm{H}_{2}\rm{O}$ and $\rm{D}_{2}\rm{O}$  are carried out to explicitly consider the energy exchange in each scattering. The results are shown in Fig.~\ref{fG_ine_zoomin}. The observed structures are diminished, confirming inelasticity is the primary origin of the observed effects. 
In other words, these patterns obtained under the elastic approximation are the signatures of inelasticity. 

\begin{table}[htbp]
    \centering
    \caption{
    The ratios of accelerated scattered neutrons, with wavelengths shorter than the incident wavelength, for the five scattering angles corresponding to Fig.~\ref{f_wl_newSim}.}
    \label{table_wl_newSim} 
    \begin{tabular}{llllll}
        \toprule
        Scattering Angle & \SI{10}{\degree} & \SI{40}{\degree} & \SI{90}{\degree} & \SI{120}{\degree} & \SI{150}{\degree}  \\  
         \midrule
        Ratio & 0.812 & 0.767 & 0.782 & 0.787 & 0.789 \\ 
        \bottomrule
    \end{tabular}
\end{table}

\section{Conclusion}
\label{sConclusion}

This study successfully reproduced the total scattering experiments of inelastic samples in the Monte Carlo simulation code, Prompt.
The simulations incorporate realistic physics, including absorption, inelasticity, and multiple scattering. The simulated results are directly comparable with experimental data, showing good agreement across various aspects, including angular and wavelength distributions, as well as angular differential cross sections.

Signatures of inelasticity, due to the shift of wavelength following neutron interactions with the samples, are also reported. 
These signatures in measurements can serve as metrics for evaluating the extent of the inelasticity effects. Moreover, the removal of these signatures is expected to be a major criterion for assessing the effectiveness of future  data correction algorithms.

Along with the capability of modelling multiple scattering, the Monte Carlo technique is expected to be feasible to correct anticipated effects in experiments. 
Further investigations will be conducted to develop a general approach for reducing and correcting experimental data guided by Monte Carlo simulations.

\section*{Acknowledgements}
We express our gratitude to Xianxiu Qiu for her assistance with sample preparation.
We would like to thank our colleagues Huaican Chen, Bo Bai, Yuanguang Xia, and Xujing Li at CSNS for their participation in the measurements.
We acknowledge the computing and data resources provided by National High Energy Physics Science Data Center, Guangdong-Hong Kong-Macao Greater Bay Area Branch (Guangdong Provincial Material Science Data Center).
The authors are grateful for valuable discussions and substantial support.

\section*{Author Contributions}
Xiao-Xiao Cai and Ni Yang led to the study conception, design and methodology.
All authors contributed to the measurements.
Simulations and data analysis were performed by Ni Yang.
Xiao-Xiao Cai, Zi-Yi Pan, Ming Tang and Ni Yang contributed to the software.
The first draft of the manuscript was written by Ni Yang, and all authors commented on previous versions of the manuscript. 

\section*{Data Availability}
A zip compressed file, including the simulated data, the data analysis scripts and an example simulation file, can be accessed through the link:
\url{https://doi.org/10.5281/zenodo.17046517}

\bibliography{reference} 

\end{document}